\documentclass[twocolumn,pre,aps,showpacs,amsmath,amssymb,superscriptaddress]{revtex4-1}

\usepackage{hyperref}
\usepackage{amsmath,amssymb}
\usepackage{amsfonts,amsthm}
\usepackage{graphics}
\usepackage{graphicx}
\usepackage{dcolumn}
\usepackage{color}
\usepackage{bm}

\begin{document}

\title{Anticipated Synchronization in a Biologically Plausible Model of Neuronal Motifs}
\author{Fernanda S. Matias\thanks{fernandasm@df.ufpe.br}}
\author{Pedro V. Carelli}
\affiliation{Departamento de F\'{\i}sica, Universidade Federal de Pernambuco,
Recife, Pernambuco 50670-901 Brazil}
\author{Claudio R. Mirasso}
\affiliation{Instituto de Fisica Interdisciplinar y Sistemas Complejos, CSIC-UIB, Campus
Universitat de les Illes Balears E-07122 Palma de Mallorca, Spain}
\author{Mauro Copelli}
\affiliation{Departamento de F\'{\i}sica, Universidade Federal de Pernambuco,
Recife, Pernambuco 50670-901 Brazil}

%\date{\today}

\begin{abstract}
  Two identical autonomous dynamical systems coupled in a master-slave
  configuration can exhibit anticipated synchronization (AS) if the
  slave also receives a delayed negative self-feedback. Recently, AS
  was shown to occur in systems of simplified neuron models, requiring
  the coupling of the neuronal membrane potential with its delayed
  value. However, this coupling has no obvious biological
  correlate. Here we propose a canonical neuronal microcircuit with
  standard chemical synapses, where the delayed inhibition is provided
  by an interneuron. In this biologically plausible scenario, a smooth
  transition from delayed synchronization (DS) to AS typically occurs
  when the inhibitory synaptic conductance is increased. The
  phenomenon is shown to be robust when model parameters are varied
  within physiological range. Since the DS-AS transition amounts to an
  inversion in the timing of the pre- and post-synaptic spikes, our
  results could have a bearing on spike-timing-dependent-plasticity
  models.
\end{abstract}

\maketitle

\section{Introduction}

Synchronization of nonlinear systems has been extensively studied on a large variety
of physical and biological systems. Synchronization
of oscillators goes back to the work by Huygens,
and in the past decades an increased interest in the
topic of synchronization of chaotic systems appeared~\cite{Boca02}.

About a decade ago, Voss~\cite{Voss00} discovered a new scheme of synchronization
that he called ``anticipated synchronization''. He found that
two identical dynamical systems coupled in a master-slave
configuration can exhibit this anticipated synchronization if the slave
is subjected to a delayed self-feedback. One of the prototypical
examples proposed by Voss~\cite{Voss00,Voss01b,Voss01a} is
described by the equations
\begin{eqnarray}
\label{eqvoss}
\dot{x} & = & f(x(t)), \\
\dot{y} & = & f(y(t)) + K[x(t)-y(t-t_d)]. \nonumber 
\end{eqnarray}
$f(x)$ is a function which defines the autonomous dynamical
system. The solution $y(t) = x(t+t_d)$, which characterizes the anticipated synchronization (AS), 
has been shown to be stable in a variety of scenarios, including
theoretical studies of autonomous chaotic
systems~\cite{Voss00,Voss01b,Voss01a}, inertial
ratchets~\cite{Kostur05}, and delayed-coupled 
maps~\cite{Masoller01}, as well as experimental observations in
lasers~\cite{Sivaprakasam01,Liu02} or electronic circuits~\cite{Ciszak09} .

More recently, AS was also shown to occur in a non-autonomous
dynamical system, with FitzHugh-Nagumo models driven by white
noise~\cite{Ciszak03,Ciszak04,Toral03b}. In these works, even when the model
neurons were tuned to the excitable regime, the slave neuron was able
to anticipate the spikes of the master neuron, working as a
predictor~\cite{Ciszak09} . Though potentially interesting for neuroscience, it is not
trivial to compare these theoretical results with real neuronal
data. The main difficulty lies in requiring that the membrane
potentials of the involved neurons be diffusively coupled. While a
master-slave coupling of the membrane potentials could in principle be
conceived by means of electrical synapses (via gap
junctions)~\cite{Kandel} or ephaptic interactions~\cite{Arvanitaki42},
no biophysical mechanism has been proposed to account for the delayed
inhibitory self-coupling of the slave membrane potential.

In the brain, the vast majority of neurons are coupled via chemical
synapses, which can be excitatory or inhibitory. In both cases, the
coupling is directional and highly nonlinear, typically requiring a
suprathreshold activation (e.g. a spike) of the pre-synaptic neuron to
trigger the release of neurotransmitters. These neurotransmitters then need to diffuse
through the synaptic cleft and bind to receptors in the membrane of
the post-synaptic neuron. Binding leads to the opening of
specific channels, allowing ionic currents to change the post-synaptic
membrane potential~\cite{Kandel}. This means that not only the
membrane potentials are not directly coupled, but the synapses themselves
are dynamical systems.

Here we propose to bridge this gap, investigating whether AS can occur
in biophysically plausible model neurons coupled via chemical
synapses. The model is described in section~\ref{model}. In
section~\ref{results} we describe our results, showing that AS can
indeed occur in ``physiological regions'' of parameter space. Finally,
section~\ref{conclusions} brings our concluding remarks and briefly
discusses the significance of our findings for neuroscience, as well
as perspectives of future work.

\section{\label{model}Model}

\subsection{Neuronal motifs}

\begin{figure}[!ht]%
\begin{flushleft}(a)%
\end{flushleft}%
\includegraphics[width=0.7\columnwidth,clip]{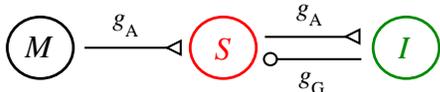}\\% 
\begin{flushleft}(b)%
\end{flushleft}
\includegraphics[width=0.7\columnwidth,clip]{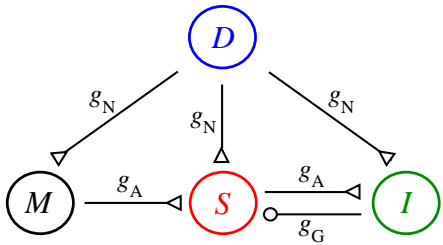}%
\caption{\label{fig:masterslave} (a) (Color online) Three neurons coupled by chemical
  synapses in the master-slave-interneuron (MSI) configuration :
  excitatory AMPA synapses (with maximal conductance $g_A$) couple
  master (M) to slave (S) and slave to interneuron (I), whereas an
  inhibitory GABA$_\text{A}$ synapse (with maximal conductance $g_G$)
  couples interneuron to slave. (b) Same as (a), except that all three
  neurons of the MSI circuit receive excitatory (NMDA) synapses from a driver neuron (D).}%
\end{figure}

We start by mimicking the original master-slave circuit of
eqs.~\eqref{eqvoss} with a unidirectional excitatory chemical synapse
(M~$\longrightarrow$~S in Fig.~\ref{fig:masterslave}a). In a scenario
with standard biophysical models, the inhibitory feedback we propose
is given by an interneuron (I) driven by the slave neuron, which
projects back an inhibitory chemical synapse to the slave neuron (see
Fig.~\ref{fig:masterslave}a). So the time-delayed negative feedback is
accounted for by chemical inhibition which impinges on the slave
neuron some time after it has spiked, simply because synapses have
characteristic time scales. Such inhibitory feedback loop is one of
the most canonical neuronal microcircuits found to play several
important roles, for instance, in the spinal cord~\cite{Shepherd}, cortex~\cite{Shepherd},
thalamus~\cite{thalamus1,thalamus2} and nuclei involved 
with song production in the bird brain~\cite{Mindlin}. For simplicity, we
will henceforth refer to the 3-neuron motif of
Fig.~\ref{fig:masterslave}a as a Master-Slave-Interneuron (MSI) system.

As we show in section~\ref{results} below, whether or not the MSI
circuit can exhibit AS depends, among other factors, on the
excitability of the three neurons. In the MSI, this is controlled by a
constant applied current (see section~\ref{sec:msi}). To test the
robustness of the results (and at the same time improve the realism
and complexity of the model), in section~\ref{sec:dmsi} we study the
four-neuron motif depicted in Fig.~\ref{fig:masterslave}b, where the
excitability of the MSI network is chemically modulated via synapses
projecting from a global driver ($D$). From now on, we refer to the
4-neuron motif as a Driver-Master-Slave-Interneuron (DMSI) microcircuit.

\subsection{Model neurons}

In the above networks, each node is described by a Hodgkin-Huxley (HH) model
neuron~\cite{HH52}, consisting of four coupled ordinary differential
equations associated to the membrane potential $V$ and the ionic currents flowing across the axonal
membrane corresponding to the Na, K and leakage currents. The gating variables for sodium are $h$ and $m$ and 
for the potassium is $n$. The equations read~\cite{Koch}:
\begin{eqnarray}
C_m \frac{dV}{dt} &=& \overline{G}_{Na} m^3 h (E_{Na}-V) + \overline{G}_{K} n^4 (E_{K}-V) \nonumber \\
&& + G_m (V_{rest}-V) + I + \sum I_{syn} \label{eq:dvdt}\\ 
\frac{dx}{dt} &=& \alpha_x(V)(1-x) -\beta_x(V) x\; , 
\end{eqnarray}
where $x\in\{h,m,n\}$, $C_m = 9\pi$ $\mu$F is the membrane capacitance
of a $30\times 30\times \pi$~$\mu$m$^2$ equipotential patch of
membrane~\cite{Koch}, $I$ is a constant current which sets the neuron excitability
and $\sum I_{syn}$ accounts for the interaction with other neurons. The reversal
potentials are $E_{Na}=115$~mV, $E_{K}=-12$~mV and $V_{rest}=10.6$~mV, which
correspond to maximal conductances $\overline{G}_{Na}=
1080\pi$~mS, $\overline{G}_{K} =324\pi$~mS and $G_m=2.7\pi$~mS,
respectively. The voltage dependent rate constants in the Hodgkin-Huxley model have the form:
\begin{eqnarray}%
\alpha_n(V) & = & \frac{10-V}{100(e^{(10-V)/10}-1)},  \\
\beta_n(V) & = & 0.125e^{-V/80}, \\
\alpha_m(V) & = & \frac{25-V}{10(e^{(25-V)/10}-1)},  \\
\beta_m(V) & = & 4e^{-V/18},\\
\alpha_h(V) & = & 0.07e^{-V/20}, \\
\beta_h(V) & = & \frac{1}{(e^{(30-V)/10}+1)}. \label{eq:betah}
\end{eqnarray}%
Note that all voltages are expressed relative to the resting potential
of the model at $I=0$~\cite{Koch}.

According to Rinzel and Miller~\cite{Rinzel80}, in the absence of
synaptic currents the only attractor of the system of
equations~\eqref{eq:dvdt}-\eqref{eq:betah} for $I \lesssim 177.13$~pA is a
stable fixed point, which loses stability via a subcritical Hopf
bifurcation at $I \simeq 276.51$~pA. For $177.13$~pA~$\lesssim I \lesssim 276.51$~pA, the stable fix point
coexists with a stable limit cycle.

\subsection{Synaptic coupling}

AMPA (A) and GABA$_\text{A}$ (G) are the fast excitatory and inhibitory synapses in
our model [see Fig.~\ref{fig:masterslave}a]. Following Destexhe et
al~\cite{KochSegev}, the fraction $r^{(i)}$ 
($i=A,G$)
of bound (i.e. open) synaptic receptors is modelled by a first-order
kinetic dynamics:
\begin{equation}
\label{eq:rate}
\frac{dr^{(i)}}{dt} = \alpha_i [T](1-r^{(i)}) - \beta_i r^{(i)},
\end{equation}
where $\alpha_i$ and $\beta_i$ are rate constants and $[T]$ is the
neurotransmitter concentration in the synaptic cleft. For simplicity,
we assume $[T]$ to be an instantaneous function of the pre-synaptic
potential $V_{pre}$:
\begin{equation}
[T](V_{pre}) = \frac{T_{max}}{1+e^{[-(V_{pre}-V_p)/K_p]}},
\end{equation}
where $T_{max}=1$~mM$^{-1}$ is the maximal value of $[T]$, $K_p=5$~mV
gives the steepness of the sigmoid and $V_p=62$~mV sets the value at
which the function is half-activated~\cite{KochSegev}.%

The  synaptic current at a given synapse is given by 
\begin{equation}
I^{(i)} = g_{i}r^{(i)}(V-E_{i}),
\end{equation}
where 
$V$ is the postsynaptic potential, $g_{i}$ the
maximal conductance and $E_i$ the reversal potential. We use
$E_{A}=60$~mV and $E_{G}=-20$~mV.

The values of the rate constants $\alpha_{A}$, $\beta_{A}$,
$\alpha_{G}$, and $\beta_{G}$ are known to depend on a number of
different factors and vary
significantly~\cite{Geiger97,Geiger02,Hausser97,Kraushaar00}. To
exemplify some of our results, we initially fix some parameters, which
are set to the values of Table~1 unless otherwise stated
(section~\ref{three}). Then we allow these parameters (as well as the
synaptic conductances) to vary within the physiological range when exploring 
different synchronization regimes (see
section~\ref{scanning} and~\ref{sec:dmsi}).

\begin{table}[b]
\begin{tabular}{|c|c|c|} 
\hline
                                  & $MSI$                     & $DMSI$\\ \hline
$\alpha_{A}$~(mM$^{-1}$ms$^{-1}$) & $1.1$  & $1.1$\\ \hline
$\beta_{A}$~(ms$^{-1}$)           & $0.19$          & $0.19$ \\ \hline
$\alpha_{G}$~(mM$^{-1}$ms$^{-1}$) & $5.0$  & $5.0$ \\ \hline
$\beta_{G}$~(ms$^{-1}$)           & $0.30$          & $0.60$ \\ \hline
$\alpha_{N}$~(mM$^{-1}$ms$^{-1}$) & ---                       & $0.072$ \\ \hline
$\beta_{N}$~(ms$^{-1}$)           & ---                       & $0.0066$ \\ \hline
$g_{A}$~(nS)                      & $10$                   & $10$ \\ \hline
 $I$~(pA)                         & $280$                  & $160$\\ \hline
             \end{tabular} 
\caption{\label{tabela:padrao}Standard values employed in the
  model. See text for details.}
\end{table}

The slow excitatory synapse is NMDA (N) and its synaptic current is given by:
\begin{equation}
I^{(N)} = g_{N}B(V)r^{(N)}(V-E_{N}),
\end{equation}
where $E_{N}=60$~mV. The dynamics of the variable $r^{(N)}$ is similar
to eq. \eqref{eq:rate} with $\alpha_{N}=0.072$~mM$^{-1}$ms$^{-1}$ and
$\beta_{N}=0.0066$~ms$^{-1}$. The magnesium block of the NMDA receptor
channel can be modeled as a function of postsynaptic voltage $V$:
\begin{equation}
B(V) = \frac{1}{1+e^{(-0.062V)[\text{Mg}^{2+}]_{o}/3.57}},
\end{equation}
where $[\text{Mg}^{2+}]_{o}= 1$~mM is the physiological extracellular
magnesium concentration.

In what follows, we will drop the neurotransmitter superscripts $A$,
$G$ and $N$ from the synaptic variables $r$ and $I$. Instead we
use double subscripts to denote the referred pre- and postsynaptic neurons. 
For instance, the synaptic current in the slave neuron
due to the interneuron (the only inhibitory synapse in our models)
will be denoted as $I_{IS}$, and so forth.

\section{\label{results}Results} 

\subsection{\label{sec:msi}Master-Slave-Interneuron circuits}

\subsubsection{\label{three}Three dynamical regimes}

Initially, we describe results for the scenario where all neurons
receive a constant current $I\geqslant 280$ pA. This corresponds to a situation
in which the fixed points
are unstable and, when isolated, they spike periodically. All other
parameters are as in Table~1. For different sets of inhibitory
conductance values $g_G$ our system can exhibit three different behaviors. To
characterize them, we define $t^{M}_{i}$ as the time the
membrane potential of the master neuron is at its maximal value in the $i$-th
cycle (i.e. its $i$-th spike time), and $t_{i}^{S}$ as the spike time
of the slave neuron which is nearest to $t_i^{M}$.

\begin{figure}[!ht]
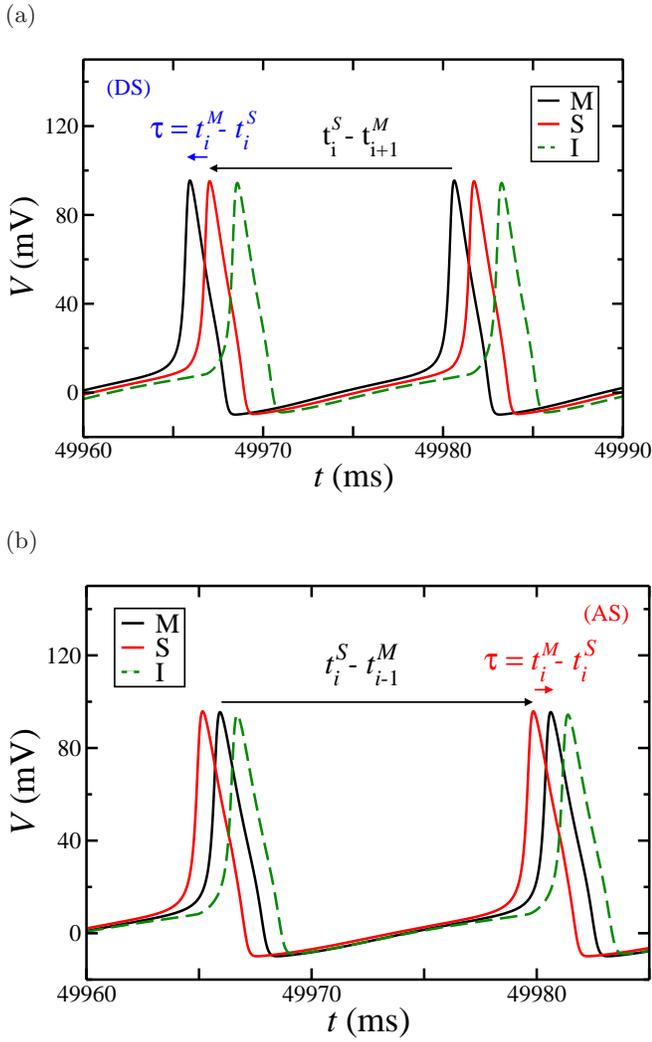
%
\begin{flushleft}(a)%
\end{flushleft}%
\centerline{\includegraphics[width=0.99\columnwidth,clip]{MatiasFig02a}}
\begin{flushleft}(b)%
\end{flushleft}%
\centerline{\includegraphics[width=0.99\columnwidth,clip]{MatiasFig02b}}
\caption{\label{fig:tau} (Color online) Membrane potential $V$ as a
  function of time for an external current $I=280$ pA in the master
  (M), slave (S), and interneuron (I) neurons. The plot illustrates two
  regimes: (a) $g_{G}=20$~nS leads to delayed synchronization (DS),
  where $\tau<0$, and (b) $g_{G}=40$~nS leads to anticipated
  synchronization (AS), where $\tau>0$. Other parameters as in
  Table~1.  }
\end{figure}

The delay $\tau_i$ is defined as the difference (see Fig.~\ref{fig:tau}):
\begin{equation}
\tau_i \equiv t^{M}_{i}-t^{S}_{i}.
\end{equation}
Initial conditions were randomly chosen for each computed time
series. When $\tau_i$ converges to a constant value $\tau$, a
phase-locked regime is reached~\cite{Strogatz}. If $\tau < 0$
(``master neuron spikes first'') we say that the system exhibits delayed
synchronization (DS)~[Fig.~\ref{fig:tau}(a)]. If $\tau>0$ (``slave neuron
spikes first''), we say that anticipated synchronization (AS)
occurs~[Fig.~\ref{fig:tau}(b)]. If $\tau$ does not converge to a fixed
value, the system is in a phase drift (PD) regime~\cite{Strogatz}. The
extent to which the AS regime can be legitimately considered
``anticipated'' in a periodic system will be discussed below.

%%%%%%%%%%%%%%%%%%%%%

\begin{figure}[!h]
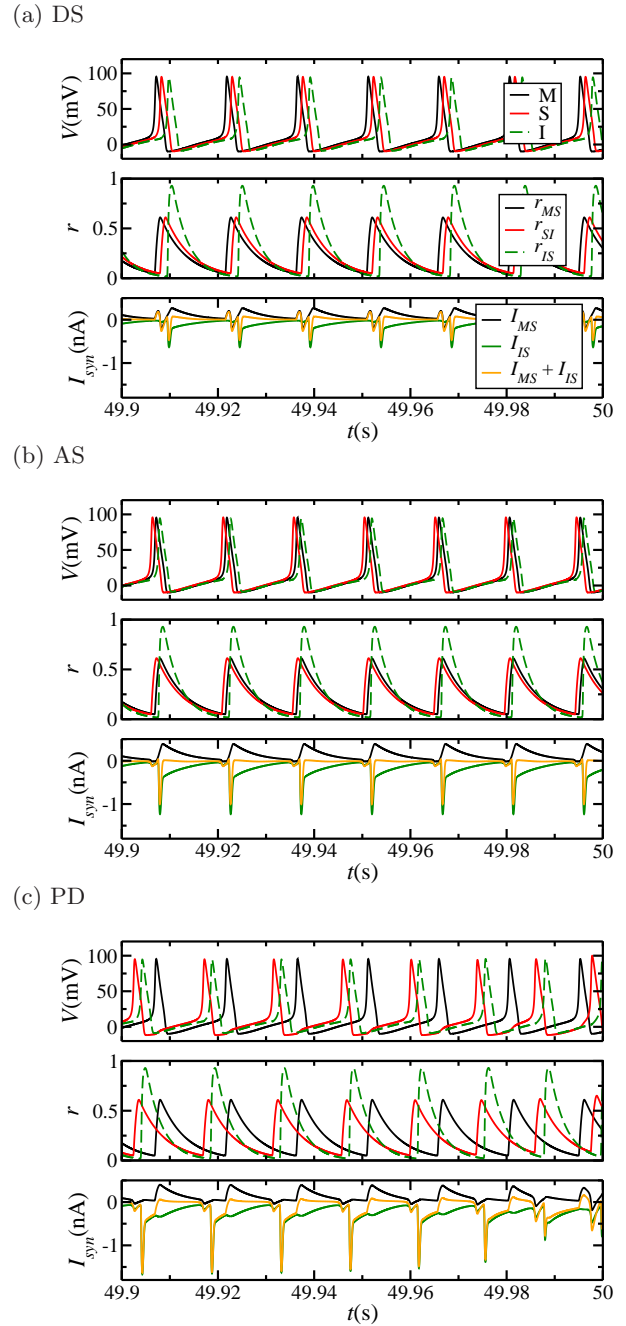
%
\begin{minipage}[t]{\linewidth}%
\begin{flushleft}(a) DS%
\end{flushleft}%
\includegraphics*[width=0.85\linewidth,clip]{MatiasFig03a}%
\end{minipage}
\begin{minipage}[t]{\linewidth}%
\begin{flushleft}(b) AS%
\end{flushleft}%
\includegraphics*[width=0.85\linewidth,clip]{MatiasFig03b}%
\end{minipage}
\begin{minipage}[t]{\linewidth}%
\begin{flushleft}(c) PD%
\end{flushleft}%
\includegraphics*[width=0.85\linewidth,clip]{MatiasFig03c}%
\end{minipage}
\caption%
{\label{fig:VrI} (Color online) Time series of the membrane potentials ($V$), bound
  receptors ($r$) and synaptic currents ($I$), with model parameters
  as in Table~1. Note that the system is periodic in the DS and AS
  regimes [(a) and (b) respectively], but not in the PD regime
  (c).}%
\end{figure}%

%%%%%%%%%%%%%%%%%%%%%%%%%%%%%%%%

In Figure~\ref{fig:VrI} we show examples of time series in the three
different regimes (DS, AS and PD). The different panels correspond to
the membrane potential, fraction of activated receptors for each
synapse, and synaptic current in the slave neuron. For a relatively
small value of the inhibitory coupling [$g_{G}=20$~nS,
Fig.~\ref{fig:VrI}(a)] the slave neuron lags behind the master,
characterizing DS. In Fig.~\ref{fig:VrI}(b), we observe that by
increasing the value of the inhibitory coupling ($g_{G}=40$~nS) we
reach an AS regime. Finally, for strong enough inhibition
[$g_{G}=60$~nS, Fig.~\ref{fig:VrI}(c)] the PD regime ensues.

\begin{figure}[!ht]%
\includegraphics*[width=0.9\columnwidth,clip]{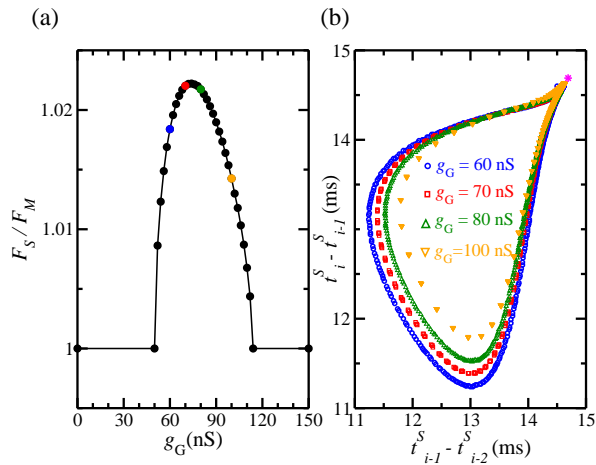}%
\caption%2
{\label{fig:firingrate} (Color online) (a) The mean firing rate of the
  slave ($F_S$) coincides with the mean firing rate of the master ($F_M$)
  for DS and AS regimes, but it is larger for PD. (b) In PD, the
  return map of the inter-spike interval of the slave is consistent
  with a quasi-periodic system (the pink star shows the return map of
  the master).}%
\end{figure}%

In the DS and AS regimes the master and slave neurons spike at the
same frequency. However, when the system
reaches the PD regime the mean firing rate of the slave neuron becomes
higher than that of the master. The counterintuitive
result shown in Fig.~\ref{fig:firingrate}(a) emerges: the mean firing rate of the
slave neuron  {\em increases\/} while increasing the conductance of the {\em
  inhibitory\/} synapse projected from the interneuron. For the
particular combination of parameters used in Fig.~\ref{fig:firingrate}(a),
the transition turns out to be reentrant, i.e., the system returns to the
DS regime for sufficiently strong inhibition (a more detailed
exploration of parameter space will be presented
below). Figure~\ref{fig:firingrate}(b) shows the return map of the
inter-spike interval of the slave, which forms a closed curve
(touching the trivial single-point return map of the master). This is
consistent with a quasi-periodic phase-drift regime.

Note that in this simple scenario $g_{G}$
plays an analogous role to that of $K$ in Eq.~\ref{eqvoss}, for which
AS is stable only when $K>K_c$ (eventually with
reentrances)~\cite{Toral03}. Moreover, the behavior of the synaptic
current in the slave neuron is particularly revealing: in the DS
regime [Fig.~\ref{fig:VrI}(a)], it has a positive peak prior to the
slave spike, which drives the firing in the slave neuron. In the AS
regime [Fig.~\ref{fig:VrI}(b)], however, there is no significant
resulting current, except when the slave neuron is already
suprathreshold. In this case, the current has essentially no effect
upon the slave dynamics. This situation is similar to the stable
anticipated solution of Eq.~\ref{eqvoss}, when the coupling term
vanishes.

\subsubsection{\label{scanning}Scanning parameter space}

\begin{figure}[!ht]%
\includegraphics*[width=0.9\linewidth,clip]{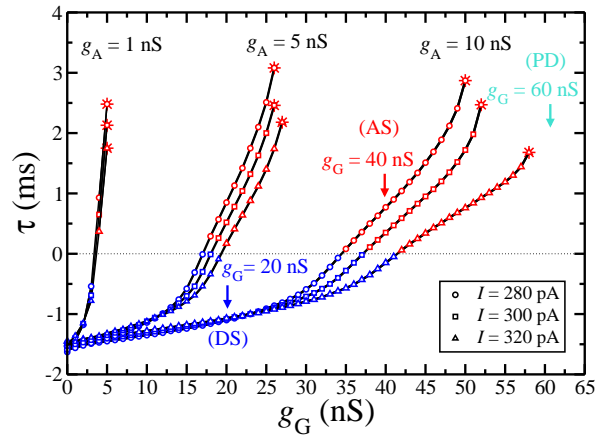}%
\caption%
{\label{fig:taugis} (Color online) Dependence of the time delay $\tau$
  with the maximal conductance $g_{G}$ for different values of the applied
  current $I$ and $g_{A}$. The end of each curve (stars) marks the
  critical value of $g_{G}$, above which the system changes from AS to
  PD.}%
\end{figure}

The dependence of the time delay $\tau$ on $g_{G}$ is shown in
Fig.~\ref{fig:taugis} for different values of the external current $I$
and maximal excitatory conductance $g_{A}$. Several features in
those curves are worth emphasizing. First, unlike previous
studies on AS, where the anticipation time was hardwired via the delay
parameter $t_d$ [see~eq.\eqref{eqvoss}], in our case the anticipation
time $\tau$ is a result of the dynamics. Note that $g_{G}$ (the
parameter varied in Fig.~\ref{fig:taugis}) does not change the time
scales of the synaptic dynamical variables ($r$), only the synaptic
strength.

Secondly, $\tau$ varies smoothly with $g_{G}$. This continuity somehow
allows us to interpret $\tau>0$ as a legitimately anticipated
regime. The reasoning is as follows. For $g_{G}=0$, we simply have a
master-slave configuration in which the two neurons spike
periodically.  Due to the excitatory coupling, the slave's spike is
always closer to the master's spike which preceded it than to the
master's spike which succeeded it [as in
e.g. Fig.~\eqref{fig:tau}(a)]. Moreover, the time difference is
approximately $1.5$~ms, which is comparable to the characteristic
times of the synapse. In that case, despite the formal ambiguity
implicit in the periodicity of the time series, the dynamical regime
is usually understood as ``delayed synchronization''. We interpret it in the following
sense: the system is phase-locked at a phase difference with a well
defined sign~\cite{Strogatz}. Increasing $g_{G}$, the time difference
between the master's and the slave's spikes eventually changes sign
[as in e.g. Fig.~\eqref{fig:tau}(b)]. Even though the ambiguity in
principle remains, there is no reason why we should not call this
regime ``anticipated synchronization'' (again a phase-locked regime,
but with a phase difference of opposite sign). In fact, we have not
found any parameter change which would take the model from the
situation in~Fig.~\eqref{fig:tau}(a) to that of
Fig.~\eqref{fig:tau}(b) by gradually {\em increasing\/} the lag of the
slave spike until it approached the next master spike. If that ever
happened, $\tau$ would change discontinuously (by its
definition). Therefore, the term ``anticipated synchronization'' by no
means implies violation of causality and should just be interpreted
with caution. As we will discuss in section~\ref{conclusions}, the
relative timing between pre- and postsynaptic neurons turns out to be
extremely relevant for real neurons.

Third, it is interesting to note that the largest anticipation time
can be longer (up to $3$~ms, i.e. about $20\%$ of the interspike
interval) than the largest time for the delayed synchronization
($\approx 1.5$~ms). If one increases $g_{G}$ further in an attempt to
obtain even larger values of $\tau$, however, the system undergoes a
bifurcation to a regime with phase drift [which marks the end of the
curves in Fig.~\ref{fig:taugis}].

\begin{figure}%[!ht]%
\includegraphics*[width=0.5\textwidth,clip]{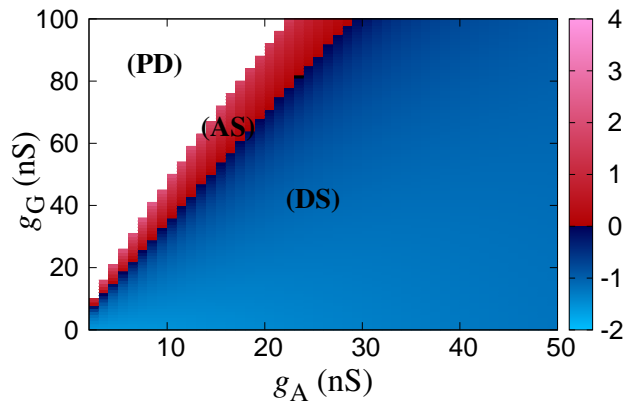}%
\caption%
{\label{fig:gegi1} (Color online) Delay $\tau$ (right bar)
  in the $(g_A,g_G)$ projection of parameter space: DS (blue, right),
  AS (red, middle) and PD (white, left --- meaning that no stationary value of $\tau$
  was found).}%
\end{figure}%

The number of parameters in our model is very large. The number of
dynamical regimes which a system of coupled nonlinear oscillators can
present is also very large, most notably $p/q$-subharmonic locking
structured in Arnold tongues~\cite{Nayfeh}. These occur in our model,
but not in the parameter region we are considering. In this context,
an attempt to map all the dynamical possibilities in parameter space
would be extremely difficult and, most important, improductive for our
purposes. We therefore focus on addressing the main question of this
paper, which is whether or not AS can be stable in a biophysically
plausible model.

In Fig.~\ref{fig:gegi1} we display a two-dimensional projection of the
phase diagram of our model. We employ the values in
Table~\ref{tabela:padrao}, except for $g_{A}$, which is varied along
the horizontal axis. Note that each black curve with circles in
Fig.~\ref{fig:taugis} corresponds to a different vertical cut of
Fig.~\ref{fig:gegi1}, along which $g_{G}$ changes. We observe that the
three different regimes are distributed in large continuous regions,
having a clear transition between them. Moreover the transition from
the DS to the AS phase can be well approximated by a
linear relation $g_{G}/g_{A}\approx 3.5$ in a large portion of the
diagram.

\begin{figure}%[!ht]%
\includegraphics*[width=0.99\columnwidth,clip]{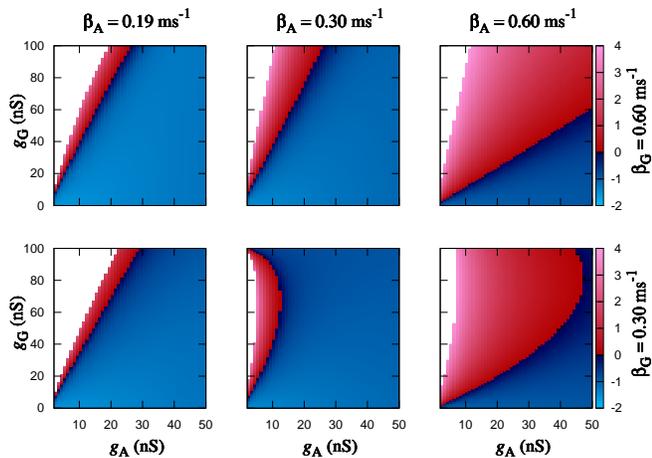}%
\caption%
{\label{fig:3betas} (Color online) Delay $\tau$ (right
  bar) in the $(g_A,g_G)$ projection of parameter space for different
  combinations of $\beta_A$ and $\beta_G$. From left to right we have respectively PD, AS and DS regimes, as in Fig.~\ref{fig:gegi1}.}%
\end{figure}%

Linearity, however, breaks down as parameters are further varied. This
can be seen e.g. in~Fig.~\ref{fig:3betas}, which displays the same
projection as Fig.~\ref{fig:gegi1}, but for different combinations of
$\beta_G$ and $\beta_A$. We observe that AS remains stable in a finite
region of the parameter space, and this region increases as excitatory
synapses become faster.

\begin{figure}%[!ht]%
\includegraphics*[width=0.5\textwidth,clip]{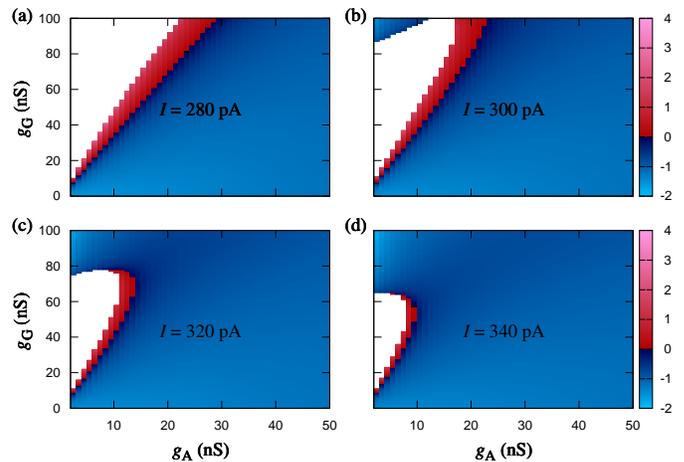}%
\caption%
{\label{fig:Ibeta} (Color online) Delay $\tau$ (right
  bar) in the $(g_A,g_G)$ projection of parameter space for different
  values of $I$. PD, AS and DS regimes as in Fig.~\ref{fig:gegi1}.}%
\end{figure}%

Figure~\ref{fig:taugis} suggests that larger values of the input
current $I$ eventually lead to a transition from AS from DS. This
effect is better depicted in Fig.~\ref{fig:Ibeta}, where the DS 
region increases in size as $I$ (and therefore the firing rate)
increases. Figures~\ref{fig:Ibeta}(b)-(d) also show that the system
can exhibit reentrant transitions as $g_{G}$ is varied. Most
importantly, however, is that Figs.~\ref{fig:3betas}
and~\ref{fig:Ibeta} show that there is always an AS region in
parameter space, as synaptic and intrinsic parameters are varied.

As we will discuss in section~\ref{conclusions}, the possibility of
controlling the transition between AS and DS is in principle extremely
appealing to the study of plasticity in neuroscience. However, in a
biological network, the input current would not be exactly constant,
but rather be modulated by other neurons. In the following, we test
the robustness of AS in this more involved scenario, therefore moving
one step ahead in biological plausibility.

\subsection{\label{sec:dmsi}Driver-Master-Slave-Interneuron circuits}

Let us consider the MSI circuit under a constant input current
$I=160$~pA. This is below the Hopf bifurcation~\cite{Rinzel80}, i.e.
none of the three neurons spikes tonically. Their activity will now be
controlled by the driver neuron (D), which projects excitatory
synapses onto the MSI circuit [see~Fig.~\ref{fig:masterslave}(b)]. We
chose to replace the constant input current by a slowly varying
current, so that the synapses projecting from the driver neuron are of
the NMDA type (see section~\ref{model}). The driver neuron receives a
current $I_D=280$~pA, so it spikes tonically. All remaining parameters
are as in the second column of Table~1.  The interest in this case is
to verify whether AS holds when the excitability of the MSI circuit is
modulated by a non-stationary current.

\begin{figure}%[!ht]%
\includegraphics*[width=0.5\textwidth,clip]{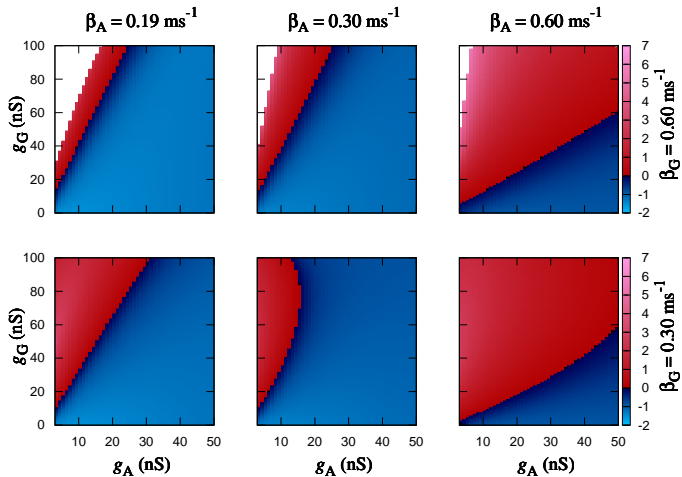}%
\caption%
{\label{fig:4betas} (Color online) DMSI circuit (see
  Fig.~\ref{fig:masterslave}b). Delay $\tau$ (right bar) in
  the $(g_A,g_G)$ projection of parameter space for different
  combinations of $\beta_A$ and $\beta_G$. PD, AS and DS regimes as in Fig.~\ref{fig:gegi1}.}%
\end{figure}

As shown in Fig.~\ref{fig:4betas}, we found in this new scenario a
similar route from DS to AS, and then the PD regime (compare with
Fig.~\ref{fig:3betas}). Note that the characteristic time ($\beta_N =
6.6$~s$^{-1}$) for the unbinding of the NMDA receptors is about ten
times larger than the inter-spike interval of the driver neuron (which
spikes at $\approx 67$~Hz). As a consequence, $r_{DM}$, $r_{DS}$,
$r_{DI}$ are kept at nearly constant values (with variations of
$\approx 10\%$ around a mean value --- data not shown). The variations
in the NMDA synaptic current are also small, which in principle should make the system
behave in an apparently similar way to the previous MSI circuit. However, these small
variations are important enough to increase the AS domain in parameter
space, in some cases even eliminating the PD region (see
e.g. Fig.~\ref{fig:4betas} for $\beta_G=0.30$~ms$^{-1}$). Therefore,
at least in this case, the use of more biological plausible parameters does not
destroy AS, but rather enhances it.

\begin{figure}%[!ht]%
\includegraphics*[width=0.5\textwidth,clip]{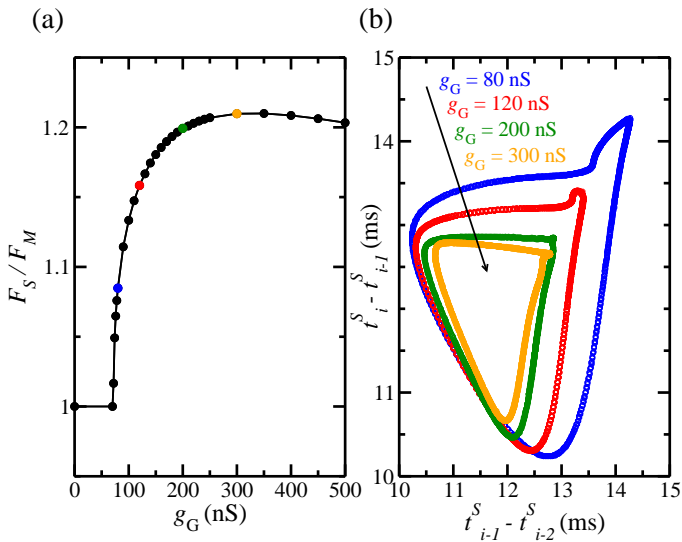}%
\caption%
{\label{fig:retornoNMDA} (Color online) DMSI circuit (see
  Fig.~\ref{fig:masterslave}b). (a) The mean firing rate of the slave
  ($F_S$) coincides with the mean firing rate of the master ($F_M$) for DS
  and AS regimes, but it is larger for PD. (b) In PD, the return map
  of the inter-spike interval of the slave is consistent with a
  quasi-periodic system.}%
\end{figure}%

In fact, the three regions in the MSI diagrams seem to retain their
main features in the DMSI circuit. When PD occurs, for example, the
slave again spikes faster than the master
[Fig.~\ref{fig:retornoNMDA}(a)], like in the MSI circuit [compare with
Fig.~\ref{fig:firingrate}(a)]. Another signature of the robustness of
the PD phase against the replacement of a constant by a slowly-varying
synaptic current appears in the return map shown in
Fig.~\ref{fig:retornoNMDA}(b). It can be seen that it has the same structure of its
three-neuron counterpart shown in Fig.~\ref{fig:firingrate}(b).

\section{\label{conclusions}Concluding remarks}

In summary, we have shown that a biologically plausible model of a
3-neuron (MSI) motif can exhibit an attractor in phase space where
anticipated synchronization is stable. The transition from the DS to
the AS regime is a smooth function of the synaptic
conductances. Typically, a further increase in the inhibitory conductance
$g_G$ leads to a second transition from AS to PD, a quasiperiodic
regime in which the slave firing frequency is larger than that of the
master.

We have varied synaptic decay rates ($\beta$), synaptic
conductances ($g$) as well as input currents ($I$) within well accepted
physiological ranges~\cite{Geiger97,Geiger02,Hausser97,Kraushaar00}.
In all the scenarios there is always a continuous region in parameter
space where AS is stable. Replacing the constant current by a global
periodic driver (arguably a more realistic situation), we obtain a
model of a 4-neuron (DMSI) motif which exhibits the same three regions
of the simpler model. The synaptic rise constants ($\alpha$) were also
varied, but have a lesser effect on the transitions among the
different regimes (data not shown). Therefore the phenomenon seems to
be robust at the microcircuit scale.

It is important to emphasize that our AS results differ from those
obtained from eq.~\eqref{eqvoss} at a fundamental level. In our model
the delayed feedback that leads to AS is given by biologically
plausible elements (an interneuron and chemical synapses). Hence, the
anticipation time is not hard-wired in the dynamical equations, but
rather emerges from the circuit dynamics. Moreover, the particular
circuit we study is a neuronal motif ubiquitously found in the
brain~\cite{Shepherd,thalamus1,thalamus2}. We are unaware of other AS
models in which every parameter has a clear biological interpretation.

%%%%%%%%%

We believe that our results can be extremely relevant for modeling studies
of synaptic plasticity. Recent decades have witnessed a growing
literature on spike-timing dependent plasticity (STDP), which accounts
for the enhancement or diminution of synaptic weight [long term potentiation 
(LTP) and long term depression (LTD), respectively] depending
on the relative timing between the spikes of the pre- and
post-synaptic neurons (see
e.g.~\cite{Abbott00,Gerstner,Clopath10}). Experimental data strongly
suggest that if the pre-synaptic neuron fires before (after) the
post-synaptic neuron, the synapse between them will be strenghtened
(weakened)~\cite{Markram97,Bi98}. STDP is supposed to take place in a
window of time differences between post- and pre-synaptic spikes in
the order of ten milliseconds, within which the delay and anticipation
times of our models fall. Since the DS-AS transition amounts to an
inversion in the timing of the pre- and post-synaptic spikes, then by
appropriately controlling this effect one could dynamically toggle
between synaptic strengthening and weakening. This could be
potentially linked with modeling of large-scale ascending feedback
modulation from reward systems.

Our results, therefore, offer a number of possibilities for further
investigation. Including effects from microcircuit dynamics (such as
the ones we have presented here) in models of synaptic plasticity is a
natural next step, one which we are currently pursuing. 
Once we have verified AS in a biologically plausible model, one could 
consider using simplified models~\cite{Neiman99,Izhikevich06} (e.g. by replacing 
the HH equations and/or the synaptic kinetics) and the influence of noise~\cite{Ciszak03,Koch}. We are also
investigating whether the structure of the phase diagram can be
qualitatively reproduced via a phase-response-curve
analysis~\cite{Ermentrout96,Rinzel98b} of the neuronal motifs studied
here. Results will be published elsewhere.

\begin{acknowledgments}
  We thank CNPq, FACEPE, CAPES and special programs PRONEX, INCeMaq
  and PRONEM for financial support. MC is grateful to the hospitality
  of the IFISC-UIB group at Palma de Mallorca, where these ideas were
  first developed. CM acknowledges support from the Ministerio de Educaci\'{o}n y Ciencia (Spain)
 and Fondo Europeo de Desarrollo Regional (FEDER) under project FIS2007-60327 (FISICOS).
\end{acknowledgments}
% 
% \pagebreak

\bibliography{FernandaSyncbib,copelli}

\end{document}